\documentclass[
 superscriptaddress,
preprint,
showpacs,
 amsmath,amssymb,
 prb,
]{revtex4-1}
\usepackage{natbib}
\usepackage{graphicx}
\usepackage[utf8]{inputenc}
\begin{document}

%Title of paper
\title{Barkhausen noise in the Random Field Ising Magnet Nd$_{2}$Fe$_{14}$B}

\author{J. Xu}
\author{D.M. Silevitch}
\affiliation{The James Franck Institute and Department of Physics, The University of Chicago, Chicago, IL 60637, USA}
\author{K.A. Dahmen}
\affiliation{Department of Physics, University of Illinois at Urbana-Champaign, Urbana, IL 61801, USA }

\author{T.F. Rosenbaum}
\email[email: ]{tfr@caltech.edu}
\affiliation{The James Franck Institute and Department of Physics, The University of Chicago, Chicago, IL 60637, USA}
\affiliation{Division of Physics, Mathematics and Astronomy, California Institute of Technology, Pasadena, CA 91125, USA}

\date{\today}

\begin{abstract}
With sintered needles aligned and a magnetic field applied transverse to its easy axis, the rare-earth ferromagnet Nd$_2$Fe$_{14}$B becomes a room-temperature realization of the Random Field Ising Model. The transverse field tunes the pinning potential of the magnetic domains in a continuous fashion. We study the magnetic domain reversal and avalanche dynamics between liquid helium and room temperatures at a series of transverse fields using a Barkhausen noise technique. The avalanche size and energy distributions follow power-law behavior with a cutoff dependent on the pinning strength dialed in by the transverse field, consistent with theoretical predictions for Barkhausen avalanches in disordered materials. A scaling analysis reveals two regimes of behavior: one at low temperature and high transverse field, where the dynamics are governed by the randomness, and the second at high temperature and low transverse field where thermal fluctuations dominate the dynamics.
\end{abstract}

\pacs{75.60.Ej,75.50.Lk,75.60.Jk,75.10.Hk}

\maketitle

\section{introduction}

The response of a ferromagnet to a time varying magnetic field has been studied extensively via experiment, numerical modeling, and theory.\cite{Barkhausen19,Laudau35,Gilbert55,Stoner48,Back99,Lakshmanan11} The magnetization, rather than evolving continuously as the field changes, often increments in discrete steps as individual domains or sets of domains rapidly switch direction. These steps were first observed as voltage pulses in inductive pickup coils wrapped around the magnet, picking up the change in magnetization with time, $\mathrm{d}M/\mathrm{d}t$. The crackling noise arising from the superposition of many such pulses is known as Barkhausen noise.\cite{Barkhausen19} Detailed measurements of the spectrum of this noise provide a sensitive probe of the internal energetics of domain reversal,\cite{Kypris14} with uses both in industrial non-destructive evaluation of magnetic materials\cite{Stefanita08} and as a valuable technique for understanding domain dynamics for model magnetic materials and Hamiltonians.\cite{Weissman88}

One important model system amenable in principle to such measurements is the Random Field Ising Model (RFIM), which is considered to be a general model for disordered systems.\cite{Sethna93} In the RFIM, the standard Ising Hamiltonian is modified by adding an applied field along the Ising axis whose value varies randomly from site to site:
\begin{equation}
H = -J\sum_{ij}\sigma_{i}^{z}\sigma_{j}^{z}+\sum_{i}h_{i}\sigma_{i}^{z}
\end{equation}
where $h_i$ is a site-random field and $J$ is the interspin interaction. The original experimental studies of the RFIM followed a proposal by Imry and Ma,\cite{Imry93} in which a site-diluted Ising antiferromagnet in a large static magnetic field forms a realization of the RFIM Hamiltonian of Eq. 1. While this approach proved fruitful for studying quantities such as the thermodynamic critical exponents\cite{Belanger83,Gofman96} and correlation lengths,\cite{Yoshizawa82} the lack of a net long wavelength moment in antiferromagnets limits the potential probes and hence the set of physical questions that can be addressed.

Within the last decade, ferromagnetic realizations of the RFIM have been discovered. They arise when diluted, dipole-coupled Ising materials are placed in a static magnetic field applied transverse to the Ising axis. The combination of the off-diagonal elements of the dipole-dipole potential and the symmetry breaking of the disorder lets the uniform transverse field generate an effective site-random longitudinal field. The initial realizations, first in the rare-earth fluoride LiHo$_{x}$Y$_{1-x}$F$_{4}$\cite{Silevitch07,Silevitch10,Tabei06,Schechter08} and subsequently in the molecular magnet Mn$_{12}$-ac,\cite{Wen10,Millis10} used dipole-coupled single spins. Both materials exhibit magnetic ordering at cryogenic temperatures due to the low energies of the dipolar coupling. Recently, Tomarken \emph{et al}. demonstrated\cite{Tomarken14} that a sintered block of the rare-earth ferromagnet Nd$_2$Fe$_{14}$B acts as a realization of the RFIM at room temperature. In Nd$_2$Fe$_{14}$B, individual spins are exchange-coupled and form elongated domains due to the inherent crystalline anisotropy.\cite{Givord84,Herbst84} These domains then interact via a dipolar coupling. The extended nature of the dipoles gives them much larger moments than individual spins would have, with a correspondingly higher energy scale (e.g. the Curie temperature for sintered Nd$_2$Fe$_{14}$B is 585 K\cite{Herbst91}). The random packing of the grains from the sintering process yields the necessary translational disorder for RFIM behavior.

We exploit the ready availability and tunability of the RFIM state in Nd$_2$Fe$_{14}$B to record the first experimental measurements of Barkhausen noise in a random-field magnet. As the transverse field deepens the pinning wells, the power law distributions in avalanche size and energy are truncated, consistent with simulation results. Critical exponents are close to mean field values in the low temperature, high transverse field regime, where we also discover extended oscillatory events in the magnetization caused by initial overshoot of the final equilibrium state by large-domain reversals.

\section{methods}

We measured the static magnetization of 2 mm diameter by 10 mm length cylinders of commercially-sintered Nd$_2$Fe$_{14}$B via Hall magnetometry, using a pair of passive GaAs Hall sensors (Toshiba THS118) in a gradiometric configuration to remove the contribution of the applied field to the Hall signal. The easy axes of the samples were oriented parallel to the cylindrical axis. We measured the domain dynamics via Barkhausen noise measurements. A 320 turn pickup coil was wrapped around the same Nd$_2$Fe$_{14}$B cylinders and an identical-geometry, empty coil wired in opposition to cancel the $\mathrm{d}H/\mathrm{d}t$ from the applied field. The signal was amplified by 10,000 with a Stanford Research SR560 low-noise voltage preamplifier, followed by a 120 kHz Krohn-Hite 3988 low-pass filter to eliminate aliasing effects, and finally a National Instruments USB-6211 250 kHz/16 bit digitizer. Longitudinal and transverse magnetic fields were supplied by a dual axis 5 T/2T superconducting magnet. Hysteresis loops were typically obtained with the transverse field fixed and the longitudinal field ramped at a constant 0.4 T/min between $\pm$ 4.5 T. Given the large field scales involved, it was not necessary to screen out the Earth's magnetic field.

For each temperature/transverse field point investigated, a series of hysteresis loops was measured to obtain a minimum of $10^{6}$ events; depending on temperature and field, this required 25 to 38 loops. For measurements at $T = 4.2$ K, helium exchange gas was used to place the sample in thermal contact with the liquid helium bath; at $T = 150$ and 300 K, the sample was in vacuum and a PID loop was used to stabilize the temperature. After acquisition, a software 1.1 kHz high-pass filter was applied to the data to remove static offsets, external power-line noise, and other spurious low-frequency effects. Individual events were identified by looking for voltages that deviated from the mean by more than 3.5 standard deviations of the time-independent background-voltage noise of the measurement chain.

\section{Results}

\begin{figure}
\includegraphics[scale = 0.5]{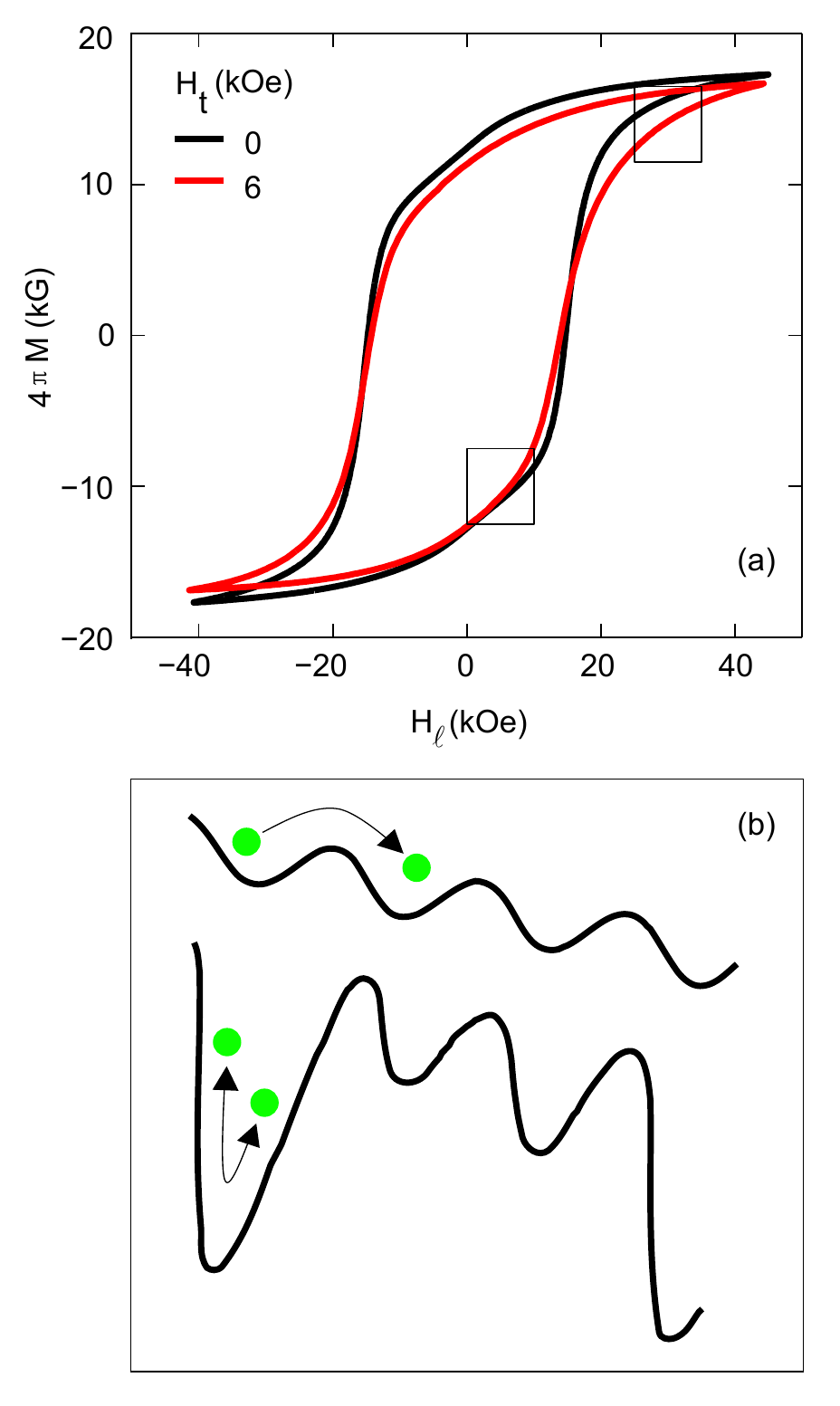}
\caption{\label{fig:cartoon}Overview of random-field effects in Nd$_{2}$Fe$_{14}$B. (a) Hysteresis loops at $T = 150$ K, with magnetization measured parallel to the Ising (easy) axis in static 0 and 6 kOe transverse magnetic fields. Application of a transverse field changes the shape of the loop, narrowing it at low longitudinal field and broadening it at high longitudinal field. The latter is an indication of enhanced pinning. Boxes at low and high longitudinal field show the ranges for the Barkhausen data plotted in Figs. \ref{fig:avalanche} and \ref{fig:oscillate}, respectively. (b) Schematic showing the evolution of the free energy landscape when a transverse field is applied. The transverse-field-induced random field tunes the system from the weak to the strong disorder limit, resulting in deeper local minima in the energy landscape, where the magnetization can oscillate about these minima.}
\end{figure}

We plot in Fig.~\ref{fig:cartoon}(a) the overall magnetic behavior of Nd$_2$Fe$_{14}$B at $T = 150$ K, well below the 585 K Curie temperature.\cite{Herbst91} Applying a constant 6 kOe transverse field results in several significant changes to the hysteresis loop. First, there is a decrease of approximately 5\% in the saturation magnetization, due to the finite value of the intrinsic anisotropy and consequent tilting of spins away from the Ising (easy) axis. Second, the shape of the loop changes when the transverse field is applied. As discussed in detail in Ref. \onlinecite{Tomarken14}, a transverse field decreases the enclosed area of the hysteresis loop for longitudinal fields below 15 kOe, while for larger longitudinal fields, applying a transverse field increases the loop area. This broadening is a signature of enhanced pinning in the high-field regime where the typical domain size is large due to coarsening, also substantiated by studies of Return-Point Memory trajectories.\cite{Sethna93,Tomarken14}

In order to study the domain reversal dynamics in the weak and strong pinning regimes, we performed Barkhausen measurements across full hysteresis loops for requisite transverse fields and temperatures. A schematic of the energy landscape is shown in Fig.~\ref{fig:cartoon}(b), where the combination of intrinsic and extrinsic pinning with the constantly ramping longitudinal field produces a tilted washboard potential for the magnetization state of the system. In the weak-pinning limit (top row), the relatively shallow minima allow the system to progress monotonically through the set of minima, resulting in a series of single-signed pulses in $\mathrm{d}M/\mathrm{d}t$ and hence single-signed spikes in the induced voltage. This is the behavior typically observed in Barkhausen experiments on magnetically soft and amorphous materials such as ferroglasses.\cite{O'Brien94,Plewka98,Zapperi98,Spasojevi96,Durin00} By contrast, increasing the depth of a number of the pinning wells (bottom row) can result in qualitatively different behavior. It becomes possible for a large reversing domain to overshoot its ultimate equilibrium point and oscillate about the center of the potential well. This results in transitory oscillatory behavior in the magnetization as the system comes to equilibrium, and hence oscillatory behavior in the pickup coil voltage. Behavior of this kind was recently reported\cite{Kypris14} in a steel alloy in which uniaxial stress was used to enhance the strength of the pinning.

\begin{figure}
\includegraphics[scale = 0.5]{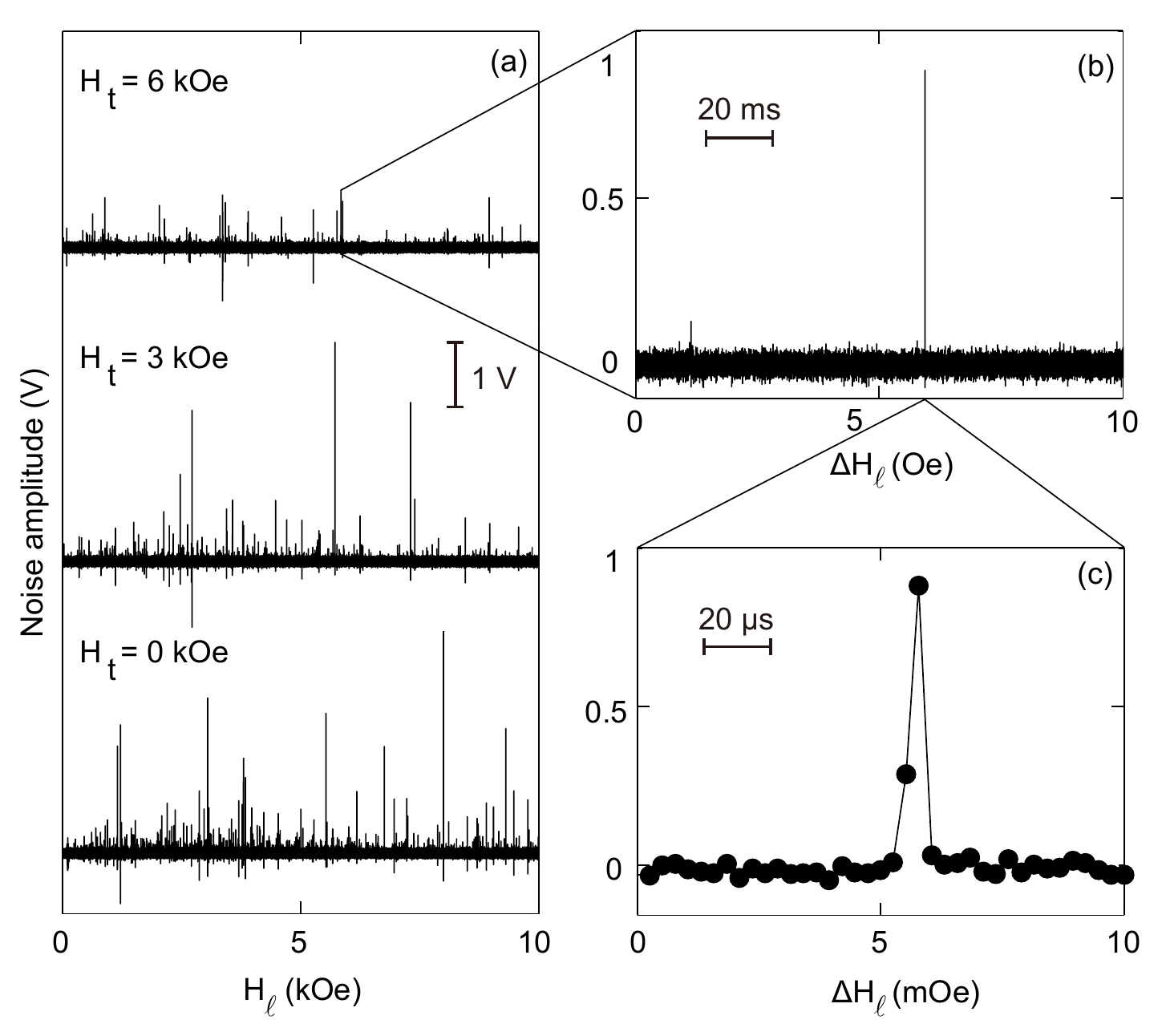}
\caption{\label{fig:avalanche}Time series of induced voltage on a pickup coil wrapped around a Nd$_{2}$Fe$_{14}$B cylinder at $T = 150$ K, low longitudinal field (increasing continuously at 0.4 T/min), and a series of transverse fields. (a) At low longitudinal fields, the absence of domain coarsening gives time series dominated by single-signed pulses, corresponding to monotonic changes in the magnetization. (b), (c) Successive enlargements of one reversal event at 6 kOe transverse field, showing the pulse structure over a duration of approximately 8 $\mu$s. }
\end{figure}

\begin{figure}
\includegraphics[scale = 0.5]{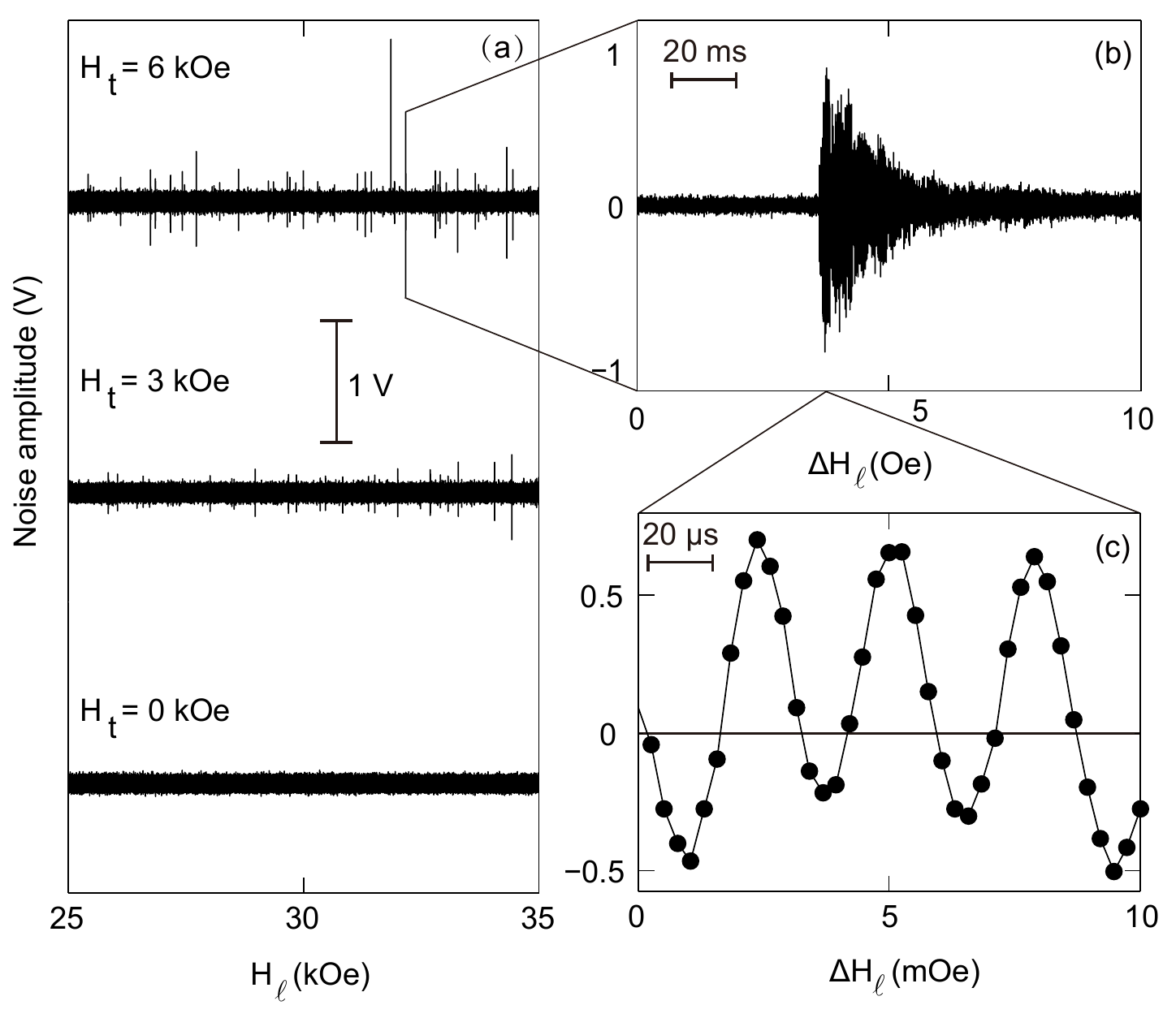}
\caption{\label{fig:oscillate}Time series of induced voltage on a pickup coil wrapped around a Nd$_{2}$Fe$_{14}$B cylinder at $T = 150$ K, large longitudinal field (increasing continuously at 0.4 T/min), and a series of transverse fields. In contrast to the behavior at low longitudinal fields (cf. Fig.~\ref{fig:avalanche}), the deep potential energy wells created by the combination of domain coarsening and the random-field-induced pinning enable magnetization oscillations. (a) Time series for a range of transverse fields. In the absence of a transverse field, the system is in the weak disorder limit and evolves in a nearly continuous fashion. As the transverse field is increased, the strength of the disorder increases, giving a steadily increasing number of discrete events. (b), (c) Successive enlargements of a single event at a longitudinal field of 3.2 T. The overall duration of the oscillatory event is approximately 30 ms, with an oscillation period of 44-48 $\mu$s (11--12 sampling points).}
\end{figure}

The salient features of our experiments are captured in Figs. \ref{fig:avalanche} and \ref{fig:oscillate}. These plots present raw time series of the voltage induced in a pickup coil surrounding a Nd$_2$Fe$_{14}$B cylinder for two segments of the hysteresis loop at $T = 150$ K (Fig.~\ref{fig:cartoon}(a)) in three fixed transverse fields. At the low longitudinal fields of Fig.~\ref{fig:avalanche}, the system consists of an ensemble of small domains that reverse monotonically as the longitudinal field is ramped, following the behavior sketched in the top row of Fig.~\ref{fig:cartoon}(b). The rate of occurrence of visible Barkhausen spikes drops as the strength of the disorder increases with increasing transverse field. This suggests that enhancing the disorder-induced pinning in this regime suppresses avalanches consisting of multiple domains reversing in quick succession to produce an observable voltage spike and forces the system towards a magnetization change dominated by small domains reversing individually, causing voltage spikes which are at or below the noise floor of our measurement chain.\cite{Sethna01} As shown in the successive enlargements of Figs.~\ref{fig:avalanche}(b) and (c), the observable events typically last less than 10 microseconds in duration. The fast dynamics are a result of the strong pinning in general in Nd$_2$Fe$_{14}$B; in the ferroglasses and other soft materials more commonly investigated using Barkhausen techniques, the reversal timescales are typically 2 to 3 orders of magnitude longer.\cite{Zapperi98,Spasojevi96,Durin00}

At the high longitudinal fields of Fig.~\ref{fig:oscillate}, we observe different signatures in the Barkhausen noise spectra. Domain coarsening effects result in large domains so that even the reversal of an individual domain can produce a detectable signal. As the strength of the pinning is again increased by ramping the transverse field, we now see an increasing frequency of events where the magnetization oscillates repeatedly, corresponding to the physical picture at the bottom of Fig.~\ref{fig:cartoon}(b). The typical oscillation period is of order a few tens of microseconds, with the ringdown period of the entire event lasting for a few tens of milliseconds (Figs.~\ref{fig:oscillate}(b) and (c)), indicating a relatively low degree of anharmonicity in the local pinning potential. We note that even in the low temperature limit the application of a transverse field is still required to induce magnetization oscillations.

\section{Discussion}

\begin{figure}
\includegraphics{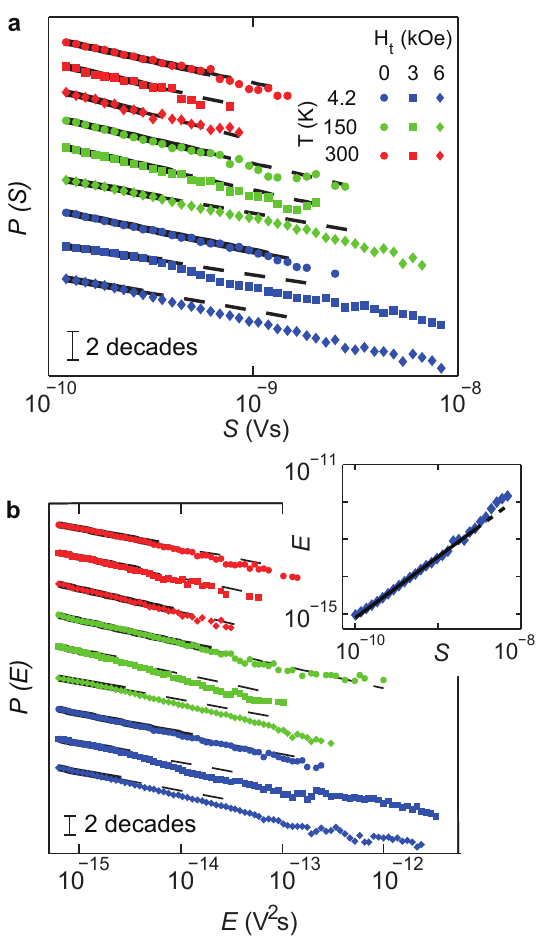}
\caption{\label{fig:prob}Probability distributions of avalanche energies and sizes for $T = 4.2$, 150, and 300 K, and transverse fields $H_t$ = 0, 3, and 6 kOe, in the low longitudinal field regime plotted in Figure~\ref{fig:avalanche}. Solid lines are fits showing the extent of pure power-law behavior; dashed lines are extrapolations of those fits, showing deviations from the power laws. Traces are offset vertically for clarity. (a) Probability distribution of integrated event area $S$. The exponents top to bottom are 3.1(0.4), 2.2(0.2), 2.5(0.3), 3.5(0.4), 3.6(0.5), 2.5(0.3), 3.5(0.4), 3.7(0.8), and 3.3(0.6). (b) Probability distribution of event energies E. The exponents top to bottom are 2.2(0.3), 2.0(0.2), 1.9(0.2), 2.3(0.2), 2.8(0.4), 1.9(0.2), 2.4(0.2), 2.3(0.7), and 2.5(0.5). Inset: avalanche size versus energy at $H_t$ = 6 kOe and $T = 4.2$ K. Solid line is a power-law fit with exponent $1.64 \pm 0.1$.}
\end{figure}

The dynamics of the monotonic switching events of the type plotted in Fig.~\ref{fig:avalanche} can be investigated by examining probability distribution histograms of different moments of the events and studying the evolution of the distributions as a function of temperature and transverse field. These probability distributions are shown in Fig.~\ref{fig:prob} for the integrated area $S = {\int(V-V_{0})}\;\mathrm{d}t$ (Fig.~\ref{fig:prob}(a)) and the total event energy $E= {\int(V-V_{0})^2}\;\mathrm{d}t$ (Fig.~\ref{fig:prob}(b)), where $V$ is the amplified time-dependent voltage and $V_0$ is the measured noise floor of the amplifier chain. $S$ is proportional to the total change in magnetization associated with a given reversal event, whereas $E$ measures the energy dissipated over the course of the event.\cite{Spasojevi96}

We observe power-law behavior for both $S$ and $E$. As a function of temperature and transverse field, the critical exponents fall into two groups. At $T = 4.2$ K and $H_{t}$ = 3 and 6 kOe, and at $T = 150$ K and $H_{t}$ = 6 kOe, the exponents are 2.3 $\pm$ 0.2 and 1.9 $\pm$ 0.2 for the distributions of $S$ and $E$, respectively. The exponents for $S$ and $E$ are comparable to the values 9/4 and 11/6 predicted from mean-field theory\cite{Perkovic99,Dahmen96}, as would be expected for a long-range interaction such as dipole-dipole coupling. Moreover, the energy distribution demonstrates a power-law relationship to the area (Fig.~\ref{fig:prob}(b) inset) with exponent $1.64 \pm 0.1$, not far from the value of 1.5 predicted from mean-field theory.\cite{Dahmen96} For all transverse fields at $T = 300$ K, as well as the low field behavior at lower temperatures, we find a different set of critical exponents: $3.5 \pm 0.4$ for $S$ and $2.4 \pm 0.3$ for $E$. This bimodal behavior suggests that we can divide the system into two regimes, the first where the disorder induced by the transverse field dominates over thermal effects and the second where the thermal behavior dominates. In the disorder-dominated regime, we see behavior broadly consistent with the mean-field expectation.

\begin{figure}
\includegraphics{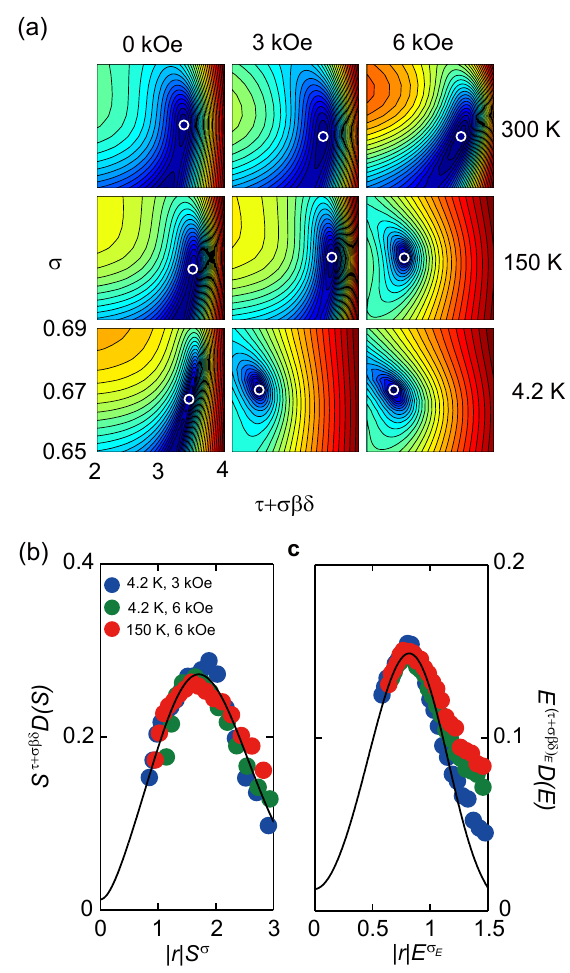}
\caption{\label{fig:scaling}Scaling behavior in the strong disorder limit. (a) ${\chi}^2$ contours for the avalanche area scaling parameters $\sigma$ and $\tau + \sigma\beta\delta$. White circles mark the locations of the best-fit minima, with deviations ranging from small (blue) to large (red). Two distinct classes of behavior are observed. At large $H_t$ and small $T$, where domain pinning is strong, there is a narrow minimum at $\sigma = 0.67 \pm 0.02$ and $\tau + \sigma\beta\delta = 2.6 \pm 0.25$. At smaller fields and/or larger temperatures, where pinning cannot compete with thermal fluctuations, qualitatively different scaling behavior is observed. There is a well-defined minimum at $\tau + \sigma\beta\delta$ = 3.75 $\pm$ 0.4 and a broader dependence on $\sigma$ with the absolute minimum remaining at $0.67 \pm 0.04$. (b), (c) Scaling collapses for the size and energy distributions in the strong disorder limit ($T = 4.2$ K, $H_t = 3$ and 6 kOe, and $T = 150$ K, $H_t = 6$ kOe). The solid lines are a phenomenological function from Ref.~\onlinecite{Perkovic95}. The scaling parameters for the size distribution (b) were $\sigma = 0.67$ and $\tau + \sigma\beta\delta = 2.53,2.47,2.70$ from top to bottom and for the energy (c) were ${\sigma}_{E} = 0.33$ and $(\tau + \sigma\beta\delta)_E = 2.2,2.1,2.1$.}
\end{figure}

We compare our measurements with the results of numerical simulations of the RFIM by examining a scaling collapse of the probability distributions of the event sizes for different temperatures and transverse fields. Following the approach of Ref.~\onlinecite{Perkovic95}, we attempt to scale the probability distributions using the functional form $D(rS^\sigma)=\lim_{R \to R_c}S^{\tau + \sigma\beta\delta}D(S,R)$, where $\sigma$ is the exponent describing the cutoff threshold, $\tau + \sigma\beta\delta$ parameterizes the slope of the probability distribution integrated around a complete hysteresis loop above the cutoff, and $r=(R - R_{c})/R$ is a reduced disorder.\cite{Perkovic95} We plot in Fig.~\ref{fig:scaling}(a)  $\chi^{2}$ contours for the scaling of all nine temperature/transverse field values as a function of the two scaling exponents. We see that the best-fit value for the cutoff threshold exponent $\sigma$ is 2/3 for all fits, with only a weak dependence on temperature and transverse field. This value exceeds the mean-field prediction of 1/2, but lies closer to it than the value of 1/4 calculated for nearest-neighbor interactions in d=3.\cite{Perkovic99} By contrast, we find that the best-fit values for $\tau+\sigma\beta\delta$ fall into two classes depending on temperature and transverse field.

In the disorder-dominated regime (transverse field large compared to temperature), the probability distributions can be scaled on top of each other with little scatter (Figs.~\ref{fig:scaling}(b) and (c)). In this regime, at $T = 4.2$ K and $H_{t}$  = 3 and 6 kOe, as well as $T = 150$ K and $H_{t}$ = 6 kOe, the area probability distributions have best-fit values of $\tau+\sigma\beta\delta = 2.6 \pm 0.25$, where as reported above $\sigma = 0.67 \pm 0.02$. This value of $\tau + \sigma\beta\delta$ compares to the mean-field expectation of 2.25 and the nearest-neighbor value of 2.03.\cite{Perkovic99} A similar scaling collapse analysis on the energy distribution for the same set of temperatures and transverse fields yields $\sigma_{E} = 0.33 \pm 0.04$ and $(\tau + \sigma\beta\delta)_{E} = 2.1 \pm 0.2$, compared to mean-field predictions of 1/3 and 11/6 respectively.\cite{Dahmen96} The probability distributions for this disorder-dominated regime can be scaled on top of each other with little scatter (Figs.~\ref{fig:scaling}(b) and (c)). By contrast, for all transverse fields at $T = 300$ K, along with 0 and 3 kOe at $T = 150$ K and 0 kOe at $T = 4.2$ K, $\tau + \sigma\beta\delta = 3.7 \pm 0.4$. In this thermally-dominated limit, the observed best-fit exponents are well away from the predictions for both mean-field and nearest-neighbor interactions. The scaling is of marginal quality, suggesting that this regime is best understood as a qualitatively different regime from the randomness-dominated regime and that a strong-disorder model is not appropriate.

Importantly, the power-law behavior does not extend to the limits of the probability distribution. As can be seen from the dashed lines in Fig.~\ref{fig:prob}, which are extrapolations of the power-law fits, the measured histograms fall off from the power law at large $S$ and $E$. This cutoff in the size and energy distributions of avalanches is a signature of a disorder-dominated system.\cite{Perkovic99,Perkovic95} For the RFIM in particular, the cutoff is expected to follow $S_{\mathrm{max}} \sim (R - R_{c})^{-1/\sigma}$, where $S_{\mathrm{max}}$ is the size of the largest avalanche for which power-law behavior is expected, $R$ is a measure of the strength of the disorder, $R_{c}$ is a critical level of disorder, and the exponent $\sigma$ is 1/2 in the mean-field limit.\cite{Perkovic99}

\begin{figure}
\includegraphics{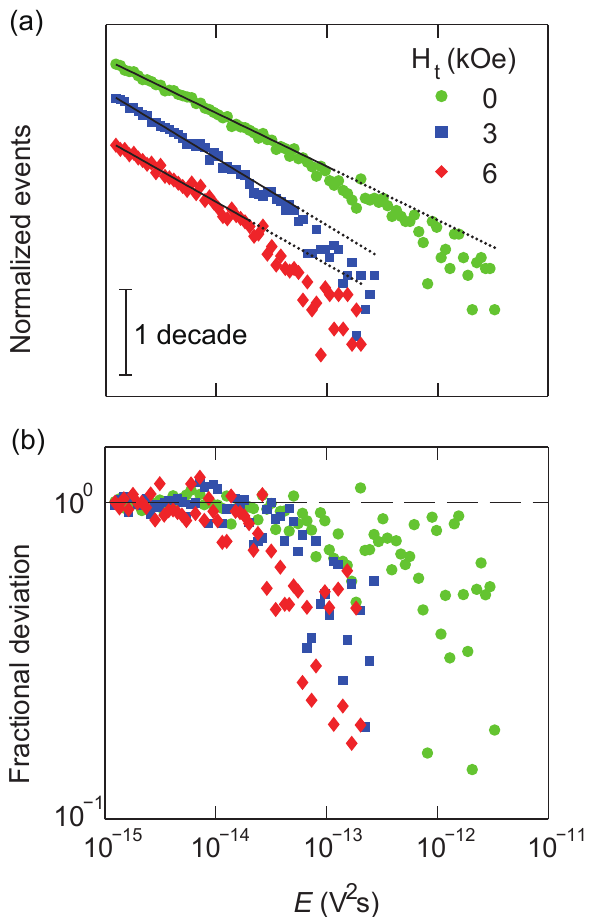}
\caption{\label{fig:depart}Departures from power-law behavior in the avalanche energy distribution. (a) T = 300 K avalanche energy distribution for a series of transverse fields fitted to a power law; traces are offset vertically for clarity. (b) Fractional deviation from power-law behavior, showing measured probability distribution divided by the individual fitted power laws.  Transverse-field-induced randomness causes systematically more profound deviations from the power-law form.}
\end{figure}

We plot in Fig.~\ref{fig:depart} the measured deviations from power-law behavior in the energy spectrum of the avalanches at $T = 300$ K for a series of transverse fields. Given that we are testing the effects of disorder, we include in this analysis avalanches from both low (Fig.~\ref{fig:avalanche}) and high (Fig.~\ref{fig:oscillate}) longitudinal fields; qualitatively similar conclusions can be drawn from considering only the single-signed Barkhausen signatures of Figs.~\ref{fig:avalanche} and \ref{fig:prob}.  In the absence of a transverse field, the disorder effects in Nd$_{2}$Fe$_{14}$B are relatively weak, and the power-law behavior extends out to the highest avalanche energies. When a transverse field is applied, the strengthening of the pinning associated with the effective random field yields a larger effective disorder, resulting in a cutoff from the power-law at systematically lower avalanche energies. This experimental result dovetails nicely with the predictions from the numerical simulations of the effects of increasing disorder.\cite{Sethna01,Perkovic99,Perkovic95}

Finally, we consider separately the magnetization reversal events prevalent at large longitudinal fields. Instead of a monotonic reversal of a domain or series of domains resulting in a single-signed peak in the induced voltage, in this regime we see events consisting of a well-defined sinusoidal oscillation which rings down over the course of 1000 or more periods (Fig.~\ref{fig:oscillate}). As the induced voltage is proportional to $\mathrm{d}M/\mathrm{d}t$, this is indicative of an oscillation in the magnetization itself. Such an oscillation requires a well-defined harmonic minimum in the free energy landscape, along with a significant degree of overshoot on the initial transition into that minimum. The first requirement suggests that the prevalence of such oscillatory events will be enhanced with increasing transverse field and hence the increasing strength of random-field-induced pinning. The large amount of inertia associated with the second requirement limits this effect to large domains reversing as a single unit, i.e. primarily in the large longitudinal field regime where coarsening has increased the typical domain size.

\begin{figure}
\includegraphics{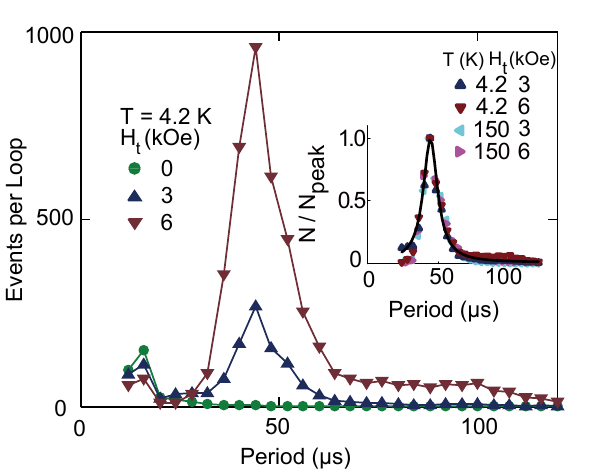}
\caption{\label{fig:periods}Distribution of periods for oscillatory magnetization events at $T = 4.2$ K and a range of transverse fields. Increasing the transverse field increases the rate of occurrence of the oscillation events, while the characteristic oscillation period remains constant. Inset: Normalized distribution of oscillation periods for $T = 4.2$ and 150 K and $H_t$ = 3 and 6 kOe, demonstrating that the shape of the distribution is essentially independent of both temperature and transverse field.  Smooth curve is a Lorentzian fit simultaneously to all four temperature/field sets.}
\end{figure}

In order to assess whether there is a single energy scale associated with these oscillatory events, we show in Fig.~\ref{fig:periods} probability distributions for the oscillation period for $T = 4.2$ K and a series of transverse fields. There are two characteristic periods for the oscillation events, with a small number of events exhibiting a $16\: \mu\mathrm{s}$ (4 pixel) period and the majority of the events showing periods ranging from 30 to above $100\: \mu\mathrm{s}$, with the predominant period at $44\: \mu\mathrm{s}$. With the application of a transverse field and stronger pinning, there is marked increase in the occurrence rate. This occurrence rate is the primary avenue for temperature and transverse-field dependence (Fig.~\ref{fig:periods} inset). The peak period and shape of the distribution curve are largely independent of the external variables, suggesting that the ratio of typical domain size in the coarsening limit to the depth of the pinning well is essentially constant and both quantities scale similarly as the disorder strength is tuned upwards via the transverse field.
A small number of short-period oscillation events are also observed at low longitudinal fields. Unlike the longer-period events in the high-longitudinal field regime, these overshoots have a typical lifespan of approximately 2 periods, presumably due to small domains oscillating in much softer harmonic wells. When a transverse field is applied and the strength of the pinning wells is increased, these short oscillation modes are progressively suppressed and monotonic reversal events dominate the probability distribution.

\section{Conclusions}

While the avalanche dynamics of the Random Field Ising Model have been studied extensively via theoretical modeling and numerical simulations, the difficulty of realizing the RFIM in bulk ferromagnets has limited the experimental possibilities. The aim of this work has been to begin to fill that gap. We have studied the reversal dynamics of the sintered rare-earth ferromagnet Nd$_{2}$Fe$_{14}$B in a transverse field, a room-temperature realization of the Random Field Ising Model. The power-law exponents and scaling forms of Barkhausen noise events as a function of temperature and transverse field demonstrate that the system can be placed into two disjoint regimes depending on tunable disorder: a strong-randomness regime in which the random-field-induced pinning dominates and a thermal regime in which fluctuations dominate. In the strong randomness regime, the critical exponents are close to mean-field predictions for heavily disordered systems. Moreover, the deep free-energy minima allow for domain-wall oscillations for extended periods of time. Having demonstrated the feasibility of using Barkhausen techniques to access the domain dynamics in this bulk random-field magnet, it should in principle be possible to extend this technique to single-spin, dipole-coupled, random-field magnets like ${\mathrm{LiHo}}_{x}{\mathrm{Y}}_{1-x}{\mathrm{F}}_{4}$ and $\mathrm{Mn}_{12}$-ac, where the random-field dynamics are enriched by the addition of quantum tunneling terms to the Hamiltonian.

\begin{acknowledgments}
Work at the University of Chicago was supported by the Department of Energy Office of Basic Energy Sciences, Grant No. DE-FG02-99ER45789.
\end{acknowledgments}

\bibliography{RandomField}

\end{document}